\documentstyle[multicol,prl,aps,epsfig]{revtex}
\begin{document}
\draft

\title{A simple physical model of liquid-glass transition: 
intrinsic fluctuating interactions and random fields 
hidden in glass-forming liquids}
\author{Hajime Tanaka}
\address{
Cavendish Laboratory, Madingley Road, Cambridge, CB3 0HE, UK. \\
Institute of Industrial Science, University of Tokyo, Minato-ku, Tokyo 106, 
Japan. \cite{A} 
}
\date{Received 17 June 1997 [cond-mat/9706174], revised 4 February 1998}
\maketitle
\begin{abstract}
We propose that glass-forming liquids are intrinsically under the 
influences of both fluctuating interactions and random fields well-known 
in the field of spin systems. This is due to 
the frustration between the isotropic and anisotropic 
parts of effective intermolecular interactions. 
Our model indicates the existence of two key temperatures relevant to 
glass transition, the density ordering point $T^\ast_{m}$ and 
the Vogel-Fulcher temperature $T_0$. 
Between $T^\ast_m$ and $T_0$, a system has features 
similar to the `Griffiths phase', while below $T_0$ it has those 
peculiar to the `spin-glass phase'.  
This picture naturally and universally explains vitrification behavior 
from its strong to fragile limit. 
\end{abstract}

\begin{multicols}{2}
Although various features of the glass transition have been clarified 
recently \cite{Angell0,Ediger,Richert,Noncry}, there has so far been no simple 
physical description of glass transition covering its strong to its 
fragile limit. 
The theoretical approaches to this problem can be grouped into 
two types: (i) one focuses on the slowing down of the dynamics on 
approach to a glass-transition temperature $T_g$.  
The topological-confinement effects on molecular motion have been 
expressed in terms of a few different concepts, 
such as free volume \cite{Cohen}, cooperative rearranging regions \cite{Adam}, 
and cage \cite{Gotze}. This stream leads to the development of 
the mode-coupling theory (MCT) \cite{Gotze}. 
However, the physical meaning of assumptions hidden in mode-coupling 
approximations is still not clear \cite{Bouchaud}. 
(ii) The other applies the knowledge of 
`spin glass' \cite{Bouchaud,Sethna2,Kirkpatrick}, 
whose glassy behavior is 
much more deeply understood than that of structural glass, 
and `frustrated systems' \cite{Kivelson,KivelsonSA} 
to the problem of glass transition. 
This type of approach is attractive in the sense that it has a 
potential to provide us with a universal physical picture 
of random frustrated systems. 
These two approaches are essentially different from each other 
in that the former presupposes disorder, while 
the latter puts more emphasis on the ordering phenomena. 
It should be noted that neither of the above approaches provides 
us with any clear answer about what physical parameter controls 
the fragility of liquids \cite{Angell0,Ediger}.   

Before making any model of glass transition, we have to seriously 
consider a much more fundamental problem, namely, {\it why some molecules 
crystallize without vitrification while others can easily form glasses 
without crystallization}: 
Without considering `crystallization', we cannot understand either a 
supercooled state, which is defined as a metastable state below 
a melting point, or vitrification phenomena, because 
the avoidance of crystallization is a prerequisite to them. 
Approach (i) regards glass transition as a purely dynamic transition. 
Thus, it usually neglects even the fact that a glass-forming system 
has the ability of crystallization. 
Approach (ii), on the other hand, fails to give a clear 
molecular-level explanation on what is the physical origin of the similarity 
between spin glass and structural glass, although it recognizes 
the importance of topological frustration 
\cite{Bouchaud,Sethna2,Kirkpatrick}. Thus, 
the wide gap between them has not been filled yet. 
Hence, neither approaches can clearly explain 
why slow dynamics starts to appear when a liquid enters 
into a metastable supercooled state upon cooling, namely, 
when a liquid is cooled below the melting point 
of its crystal. 

The main aim of this Letter is to clarify the physical origin 
of frustration in glass-forming liquids that apparently have no intrinsic 
quenched disorder and to establish a direct 
connection between structural glass and spin glass on the basis 
of molecular-level consideration. 
In contrast to the common sense that density $\rho$ is the only 
order parameter required for the physical description of liquids, 
we propose that the introduction is necessary of a new order parameter (bond 
parameter) representing a 
locally favored structure in liquids: 
This local bond formation 
causes `fluctuations in the intermolecular interactions' and 
`symmetry-breaking random fields' in much the same way 
as in spin systems \cite{Bouchaud,Dotzenko}. 
On the basis of the knowledge of spin systems \cite{Dotzenko}, 
we propose a universal picture of glass transition. 
To our knowledge, this is the first approach to 
the problem of the liquid-glass transition directly 
focusing on `crystallization (or density ordering)'.  

First we focus on the effective attractive interaction potential between a 
molecule and its  neighboring molecules. 
It is generally given by 
the form $V(r,\Omega) =\bar{V}(r)+\Delta V(r,\Omega)$, 
where $r$ is the distance from the center of mass of 
the molecule and $\Omega$ expresses the orientation. 
$\bar{V}$ represents the isotropic part of the interaction 
and $\Delta V$ its anisotropic part.  
This anisotropy can lead to a locally favored 
structure made of a molecule and its neighboring $n$ molecules, which 
is not consistent with the crystallographic symmetry. 
Thus, there can exist competing interactions in any liquids. 
Even for spherical-particle systems, it is known that  
an `icosahedral structure' is locally favored, which is separated 
from alternative arrangements such as bcc and fcc 
by high potential energy barriers ($\sim n \Delta V$, $n=12$). 
Its importance in the glass problem was widely recognized by many researchers 
\cite{Sethna2,Kirkpatrick,Kivelson,KivelsonSA,Frank,LORO,Kawasaki}. 
This local bond ordering is incompatible with any long-range 
order, so that it plays the role of a random disordering field against 
crystallization: it favors vitrification. 
This competition between density and bond ordering results from 
two conflicting requirements: (i) to minimize the distance between 
nearest-neighbor molecules and (ii) to maximize the number of surrounding 
molecules. 
In strong liquids, on the other hand, the locally favored symmetry 
is mainly selected 
by specific anisotropic interactions between molecules \cite{HTJCP} 
such as hydrogen and covalent bonding, 
which often lead to a `tetrahedral structure'. 
For a general physical description of 
real liquids, thus, we need a new order parameter $\mbox{\boldmath$S$}$ 
to describe the presence of such locally preferred arrangement 
of liquid molecules:  
$\rho$ and $\mbox{\boldmath$S$}$ are the minimal order 
parameters required for the physical description of the above 
complex features of many-body interactions.  
Although $\mbox{\boldmath$S$}$ should have a tensorial 
character that plays important roles in  the selection 
of crystallographic symmetry and its rotational dynamics, 
we can treat it as a scalar order parameter $S$ when we consider 
the phase behavior. The relevance of such an approximation is well 
established in the field of liquid crystals \cite{Lubensky}. 
Without loosing generality, therefore, 
the bond parameter $S$ can be defined as 
the `local number density of locally favored structures':
$S(\vec{r})=\Sigma_i \delta (\vec{r}-\vec{r}_i)$, 
where $\vec{r}_i$ is the position vector 
of a locally favored structure (number $i$) and 
$\Sigma_i$ is the sum about $i$ over a unit volume.  
Note that the spatial distribution of  
locally favored structures ($\vec{r}_i$) is 
random due to the nature of bond formation \cite{random}. 
The spatially averaged value of $S$ is given by 
$\bar{S}=S_0 \exp(\beta n \Delta V)$, where $\beta=1/k_{\rm B} T$ 
($k_{\rm B}$ is the Boltzmann constant and $T$ is the temperature), since 
a locally favored structure, which is stabilized by $n$ bonds, 
is in a lower energy state than the other part of liquid 
by $n \Delta V$. 
When the dependence of $\Delta V$ on $\Omega$ is not consistent with the 
crystallographic symmetry, 
$S$ decreases the local density, lowers 
the crystallization temperature and thus 
locally disturbs crystallization. 
Thus, each molecule itself internally has the cause of 
disorder and random fields against the density ordering.  
We believe that this frustration between $\rho$ and $S$ is the 
physical factor determining how easily molecules can form a glass 
without crystallization. 
Note that the strength of disorder effects  
increases with $\Delta V$, and so 
the stronger glass suffers from the stronger spin-glass (SG) effects. 

The above consideration leads to the following two-order-parameter 
description of liquids: In contrast to the common sense that a 
liquid state can be described by a single order parameter, $\rho$, 
we need a new order parameter $S$ to express a tendency of 
local bond ordering since 
in any liquids molecules locally favor a certain symmetry, 
which is not necessarily consistent with a crystallographic 
symmetry favored by by density ordering. 
This competition between the two order parameters, 
$\rho$ and $S$, 
causes frustration, which we believe is the major origin of vitrification. 
This physical picture can be naturally modeled by a 
Ginzburg-Landau-type model with the couplings between 
$\rho$ and $S$, which represent disorder 
effects of $S$ on the ordering of $\rho$. 
Density fluctuations $\delta \rho$ in the liquid phase indicating 
the instability toward the solid phase have a maximum 
at nonzero wavenumber $q_0$, whose essential feature 
is well characterized by the following bare static structure factor: 
$S_0(q)=2k_BT\chi_0(q)=k_BT/[\tau+K(q-q_0)^2]$. 
The phenomenological free energy which predicts the above $S_0(q)$ 
is given by \cite{Lubensky} 
\begin{eqnarray}
\beta H_{\rho}&=&\int d\vec{r} d \vec{r'} \delta \rho(\vec{r}) 
\chi_0^{-1}(\vec{r}-\vec{r'}) 
\delta \rho (\vec{r'})- \frac{f}{3}\int d\vec{r} \delta \rho(\vec{r})^3 
\nonumber \\
&+&\frac{g}{4}\int d\vec{r} \delta \rho(\vec{r})^4, 
\end{eqnarray}
where $\tau=a_2(T-T^\ast_\rho)$ and $T^\ast_\rho$ is  
the temperature of the mean-field limit of stability of the liquid phase. 
This Hamiltonian implies that `density order parameter 
favors spherical symmetry'. 
By including the coupling between 
$\delta \rho$ and $S$  
into the above standard theory of a liquid-solid transition, 
we obtain the following Hamiltonian that we believe is relevant to the 
glass transition: 
\begin{eqnarray}
\beta H_{GT}=\beta H_{\rho}+\int d\vec{r}\ 
[-c_1 \delta \rho(\vec{r}) S(\vec{r}) 
- \frac{c_2}{2} \delta \rho(\vec{r})^2 S(\vec{r})].  
\label{HGT}
\end{eqnarray}   
In the above, the coupling between $\rho$ and $S$ 
is introduced through the coupling constants $c_i$. 
For the ``negative'' coupling ($c_i<0$), the formation of 
active bonds (or $S$) leads to a decrease of local density and 
also to a decrease of the ordering temperature. 
For the ``positive'' coupling ($c_i>0$), on the other hand, 
the formation of active bonds (or $S$) leads to an increase in 
local density and to an increase of the ordering temperature, and 
so molecules should never form a glass and just crystallize. 
{\it The type (sign) of coupling between $\rho$ and $S$ gives 
a simple criterion on whether molecules 
crystallize without vitrification or can easily form a glass.} 
Hereafter we consider the case of $c_i<0$, since vitrification 
can occur only for this case. 

The dynamics of $\delta \rho$ can be described by \cite{Kirkpatrick}
\begin{eqnarray}
\frac{\partial \delta \rho(\vec{r},t)}{\partial t}=
\Gamma_0 \nabla^2 \frac{\delta (\beta H_{GT})}
{\delta (\delta \rho(\vec{r},t))} +\zeta(\vec{r},t), 
\end{eqnarray}
where $\zeta$ is the noise term and $\Gamma_0$ is a bare kinetic 
coefficient. 
Although $S$ is not a quenched variable, the much slower dynamics of $S$ 
than $\rho$ guarantees the quasi-quenched nature of $S$: 
The lifetime of locally favored structures is longer 
than the characteristic time of density fluctuations by at least 
a factor of $R \sim \exp(\beta n \Delta V)$.  
More importantly, recent theoretical studies indicate that even a 
frustrated random system without quenched disorder has essentially the same 
features as that with quenched disorder \cite{Bouchaud}. 
These facts allow us to regard $S$ as a quenched variable 
as long as we consider disorder effects of $S$ on density ordering. 
Thus, we treat $S$ as a quenched variable hereafter.  

In relation to the above, we briefly consider the stability of 
a metastable supercooled state. 
First we define $T_m^\ast(0)=T_m^\ast$ and $T_m^\ast(\bar{S})$ respectively 
as the density-ordering temperature 
(a crystallization or melting temperature) without and 
with the effects of random disorder 
($\bar{S}$ is the disorder strength). 
The former is the melting temperature of a 
real defect-free crystal formed as a result of density ordering, 
while the latter is that of a hypothetical crystal with 
quenched disorder (or local bond ordering), 
which may not exist in reality. 
The melting temperature of this hypothetical crystal with 
defects, $T_m^\ast(\bar{S})$, rapidly decreases with 
an increase in defect density, or $\bar{S}$ 
(see figure \ref{fig:PD}), which is known as `dilution effects' 
in spin systems. 
For real crystallization, which 
requires the annealing of the random nature of $S$, on the other hand, 
a system has to overcome a large energy barrier corresponding to 
the break-up or deformation of all the active bonds 
in a critical nucleus of crystal, which costs 
the energy of the order of $\Delta E \sim n \Delta V \bar{S} v_n$, 
where $v_n$ is the volume of a critical nucleus. 
This naturally explains why a metastable supercooled state is so 
stable in glass-forming liquids. 
Crystallization at $T_m^\ast$ can thus be kinetically avoided 
by this extra energetic barrier for nucleation, $\Delta E$, 
for a sufficient cooling rate and thus 
vitrification can be induced, as described below. 
In the metastable branch of a supercooled liquid, therefore, 
the random distribution of locally favored structures is not altered 
\cite{random} by density ordering . 
Provided that crystallization is 
kinetically prohibited, we can regard a supercooled state 
as the quasi-equilibrium thermodynamic state.  
Thus, we consider a problem of vitrification on the basis 
of this quasi-equilibrium assumption. 

Here we point out the similarity of the above Hamiltonian 
$H_{GT}$ [see equation (\ref{HGT})] assuming a quenched nature of $S$ 
and that of a spin system under fluctuating 
interactions and random fields. The Ginzburg-Landau-type Hamiltonian 
under random transition temperatures $\delta \tau(\vec{r})$, $H_{QD}$, 
and that under random fields $h(\vec{r})$, $H_{RF}$, are \cite{Dotzenko}: 
\begin{eqnarray} 
\beta H_{QD}&=&\int d\vec{r}\ [\frac{1}{2}K(\nabla \phi)^2+\frac{1}{2}
(\tau-\delta \tau(\vec{r}))\phi^2+\frac{1}{4}g \phi^4], \nonumber \\
\beta H_{RF}&=&\int d\vec{r}\ [\frac{1}{2}K(\nabla \phi)^2+\frac{1}{2}
\tau \phi^2+\frac{1}{4}g \phi^4 +h(\vec{r}) \phi], \nonumber
\end{eqnarray} 
where $\phi(\vec{r})$ is the order parameter.  
Our Hamiltonian $H_{GT}$ has both features of $H_{QD}$ and $H_{RF}$ 
at the same time: 
By setting $\phi(\vec{r})=\delta \rho(\vec{r})$, 
$\delta \tau(\vec{r})=c_2 S(\vec{r})$, and $h(\vec{r})=-c_1 S(\vec{r})$, 
$H_{GT}$ can be directly correlated to $H_{QD}$ and $H_{RF}$, 
after coarse graining the high $q$ mode around $q_0$. 
Thus, {\it the essential effects of 
random disorder on density ordering in our system 
should be the same as those in spin systems} \cite{Bouchaud,Dotzenko}. 
The quasi-quenched nature of $S$ allows us to consider the problem 
in terms of complex energy landscape, without solving the dynamic 
equations. 

On the basis of the knowledge of spin glass \cite{Bouchaud,Dotzenko}, 
we draw a simple physical picture of glass transition 
(see figure \ref{fig:PD}). 
Above $T_m^\ast$, the system behaves just as an ordinary liquid and 
the structural relaxation time $\tau_\alpha$ obeys a simple Arrhenius law, 
$\tau_\alpha=\tau_\alpha^\infty \exp(\Delta G/k_BT)$, where $\Delta G$ is 
the activation energy and an increasing function of $\Delta V$, 
and the relaxation function $\Phi(t)$ is exponential: 
$\Phi(t)=\Phi(0) \exp(-t/\tau_\alpha)$. 
Below $T_m^\ast$, 
the saddle-point equation $\delta H_{GT}/\delta (\delta \rho)=0$, 
namely, $-K\nabla^2 \delta \rho(\vec{r})+[\tau-c_2S(\vec{r})]\delta 
\rho(\vec{r}) +g\delta \rho(\vec{r})^3-c_2 S(\vec{r})=0$, starts 
to have local minimum solutions around the high-density side 
of the liquid free-energy minimum \cite{note}.
This means that there exist a macroscopic number 
of spatial `islands' having a higher density than 
the liquid, below $T_m^\ast$. 
This situation is similar to the spin glass where the phase 
space is factorized into a hierarchy of `valleys', 
or pure states of local minima separated by macroscopic barriers. 
Thus, slow relaxations overcoming barriers separated by different valleys 
are expected. 
This phase existing between $T_m^\ast$ and $T_m^\ast(\bar{S})$, 
which is characterized by the existence of numerous metastable states 
separated by finite barriers, is similar to the ``Griffiths phase''  
\cite{Griffiths,Bray} known in spin systems. 

This `Griffiths-like phase' can be characterized as follows 
\cite{Dotzenko,Griffiths,Bray}. 
In the temperature interval $T_m^\ast(\bar{S})<T<T_m^\ast$, 
$\Phi(t)$ is described by the stretched exponential 
\cite{Bray,Palmer}: 
$\Phi(t)=\Phi(0) \exp[-(t/\tau_\alpha)^{\beta_K}]$, 
instead of the usual exponential relaxation, 
as it should be in the ordinary liquid phase (the paramagnetic 
phase in spin systems) \cite{note2}. 
In spin systems \cite{Ogielski}, it is claimed that the stretched 
exponential parameter $\beta_K$ is the temperature-dependent exponent, as it 
is a finite value ($<1$) at $T=T_m^\ast(\bar{S})$, and it increases 
monotonically up to $\beta_K=1$ at $T_m^\ast$. 
This is consistent with what has been observed in structural glasses 
\cite{Richert,Noncry,note2}. 
Further, the strong coupling between 
`islands' probably leads to the existence of the SG-like phase 
below $T_{SG}$, in which the numerous disorder-dependent local minima 
are probably separated by macroscopic (or infinite) energy barriers 
\cite{Dotzenko}. 
We assign this phase transition at $T_{SG}$ from the Griffiths-like phase to 
the SG-like phase, to the Vogel-Fulcher temperature $T_0$, 
where the relaxation time diverges due to the infinite barriers 
and the ergodic to non-ergodic transition takes place. 
However, this point is never reached in real 
experiments because the large barrier heights near $T_0$ 
cause extremely slow relaxations even above $T_0$. 
Then, the glass-transition temperature $T_g$ can be defined as 
a temperature where the metastable 
`islands' having sufficiently high energetic barriers do percolate.
Thus, $T^\ast_m>T_g>T_0>T^\ast_m(\bar{S})$. 

Figure \ref{fig:PD} shows a phase diagram 
of glass transition on the basis of the above physical picture.  
Here it should be noted that for the liquid-glass transition there can be two 
types of the origins of disorder whose temperature dependencies are 
crucially different from each other: 
(i) anisotropic interactions that are not consistent with the 
crystallographic symmetry, in molecular glass formers ; and  
(ii) quenched disorder in structures of particle or molecules, 
{\it e.g.} the polydispersity of colloid 
particle sizes and the disorder in the stereoregularity, tacticity, 
and chemical structures of polymers. 
For case (i) the strength of disorder and random fields ($\bar{S}$) 
is strongly dependent upon $T$, while for case (ii) 
the strength of the disorder is independent of $T$. 
The temperature-cooling paths 
drawn in figure \ref{fig:PD} correspond to case (i).  

Here we consider the non-Arrhenius 
behavior below $T_m^\ast$ in more detail. 
We assume the divergence of the barrier height $E_{barrier}$ 
between `islands' is given by $E_{barrier}=BT/(T-T_0)$. 
This gives us the well-known Vogel-Fulcher (VF) law.  
Here we do not repeat the argument deriving this relation 
(see {\it e.g.} \cite{Sethna2,Kirkpatrick,Dotzenko,Palmer,FisherDS} 
on the details). 
Instead, we stress that the effective barrier 
dominating the structural relaxation $E_{barrier}^{eff}$ 
should be given by (i) $E_{barrier}^{eff}=\Delta G$ above $T_m^\ast$ and (ii)  
$E_{barrier}^{eff}=\Delta G+BT/(T-T_0)$ below $T_m^\ast$. 
This can be expressed analytically by the following 
modified VF law:                              
\begin{equation}
\tau_\alpha=\tau_\alpha^\infty \exp(\Delta G/k_BT) 
\exp [f(T) B/k_B(T-T_0)]. \label{MVF}
\end{equation}
Here $f(T)$ is  the Fermi-distribution-like function: 
$f(T)=1/(\exp(\gamma (T-T_m^\ast))+1)$, 
where $\gamma$ is a positive constant that controls the sharpness of the 
transition at $T^{\ast}_m$. 
This modified VF law can naturally describe the crossover from 
the Arrhenius to non-Arrhenius behavior at $T_m^\ast$ and the divergence 
at $T_0$. 
The fitting of the traditional VF law to data in a low temperature 
range often produces an unrealistic attempt frequency 
$\tau_\alpha^\infty$ (see {\it e.g.} \cite{Fischer2}). 
This problem can be removed by using equation (\ref{MVF}).
From the condition of the disappearance of the non-Arrhenius behavior 
above $T^\ast_m$, further, we obtain the relation 
$B \sim \Delta G (T^\ast_m-T_0)/T^\ast_m$. The 
fragility parameter $[D]$ estimated by this relation with $[D]=B/T_0$ 
is well correlated 
with $D$ determined experimentally, as shown in the inset of 
figure \ref{fig:TmT0}. 
This provides us with a clear physical reasoning on 
the Angell plot \cite{Angell}, including the correlation between 
bond strength $S$ and the strong nature of liquids.  
  
Finally, we check the other main predictions of our model: 
(i) there occurs a dynamic transition from the non-cooperative 
(Arrhenius-type) to cooperative regime (Vogel-Fulcher-type 
behavior) at $T_m^\ast$ (near $T_m$); 
(ii) the disorder strength $\bar{S}$, which can be correlated with 
the activation energy 
$\Delta G$ of the $\alpha$ relaxation above $T_m^\ast$, determines 
the temperature distance between $T_m^\ast$ and $T_0$;  
(iii) a stronger glass suffers from stronger disorder effects; 
(iv) an increase in pressure increases $\bar{\rho}$, but 
decreases $\bar{S}$ for $c_i<0$, simply because 
a locally favored structure having a greater volume is destroyed 
by applying pressure.  
This weakens the random disorder effects, which makes the system more 
fragile. 
Prediction (i) is quite natural and has already been recognized 
by many researchers (see e.g. \cite{Fischer2}), 
although there has so far been no physical 
reasoning on it. We stress that none of previous models 
can explain this fact, since no models of glass transition 
have so far focused on crystallization phenomena themselves. 
This prediction is quite specific in the sense that $T^\ast_m$ 
is a measurable quantity, while $T_c$ in MCT and 
$T^\ast$ in the theory of Kivelson et al. are not.  
Prediction (ii), on the other hand, can be confirmed by 
figure \ref{fig:TmT0}, which indicates the strong correlation 
between $\Delta G$ (or $\Delta V$) and $T_m-T_0$. 
Prediction (iii) is consistent with : (a) the fact that the change in 
specific heat 
across $T_g$ is weaker for a stronger glass (see {\it e.g.} 
figure 4 of \cite{Angell0}); (b) the recent finding 
of Sokolov et al. \cite{Sokolov2} on the stronger phonon scattering 
in stronger liquids; (c) a larger distance between $T_g$ 
and $T_0$; and (d) the resulting larger $\beta_K$ at $T_g$ 
for a stronger glass (see {\it e.g.} \cite{Richert}). 
Prediction (iv) is also quite consistent with the experimental 
findings by Cook et al. \cite{Angell0,Cook} that 
strong liquids become more fragile under a high pressure. 
This cannot be explained by the theory like MCT 
that is described by a single order parameter (density), since 
a plot of viscosity versus density should be universal in such models. 
The relevance of our two-order-parameter (spin-glass) model to 
the glass transition is strongly 
supported by the validity of these predictions  
and also by the experimental results suggesting the similarity 
between structural glass and spin glass \cite{Leheny}. 

It is worth noting here the relation between our model 
and the model of Kivelson et al. \cite{Kivelson,KivelsonSA}, since 
both put a focus on `frustration'. 
The most crucial difference is that we take a crystalline state 
(the ordering temperature $T^\ast_m$) as the reference state, while 
they take a `postulated' quasicrystal of a locally favored 
structure (the ordering temperature $T^\ast$), 
which is prohibited in a real system due to frustration.  
In other words, they presuppose the avoidance 
of crystallization by hand. This leads to essential 
differences in physics, although both temperatures are claimed 
to be located near $T_m$. 
The correlation of $T^\ast_m$ to $T_m$ is quite natural 
for our case, while there seems to be no such justification 
for their case since we cannot expect any correlation 
between the hidden ordering point of a frustrated 
quasicrystal and that of a real crystal, which have 
essentially different symmetries. 

In summary, we propose a simple physical picture of glass 
transition on the basis of `the two-order-parameter description 
of liquids' \cite{HTunp}, which connects structural glass and spin glass 
in a natural way. 
In our view, vitrification is a result of the competition 
between density ordering and hidden local bond ordering. 
Our study indicates that frustrated systems such as spin glass 
and structural glass have the universal phase and dynamic behavior 
characterized by a complex energy landscape 
peculiar to `Griffiths-like' and `spin-glass' phases. 
Here it should be noted that the fragile limit 
($S \rightarrow 0$) with $c_2=0$ of the dynamic version of our model 
is mathematically identical to the pioneering theory of 
Kirkpatrick and Thirumalai 
\cite{Kirkpatrick}, which further has dynamic features 
similar to MCT  
\cite{Gotze} (see the discussions in \cite{Kirkpatrick}). 
Thus, our model may give a clue to solve the problem of the existence of 
hidden disorder in mode-coupling equations pointed out by Bouchaud et al. 
\cite{Bouchaud}. 
Finally, we stress that our model can provide us with {\it a universal 
description of the glass transition covering its strong to fragile limit}. 
The strong nature of liquids increases with an increase in 
the disorder strength against density ordering ($c_1S$ and $c_2S$). 

The author is grateful to S F Edwards, E M Terentjev, C -Y D Lu, 
and D V Grinev for valuable discussions and comments. 
He also thanks E W Fischer, S Dietrich, R Evans, M Mezard, and S Franz 
for enlightening discussions. He acknowledges 
a fruitful stay at Cavendish Laboratory. 
His stay was made possible by the financial support  
from the Ministry of Education, Science, Culture, and Sports, Japan. 


\begin{minipage}{8cm}

\begin{figure}
\begin{center}
\psfig{figure=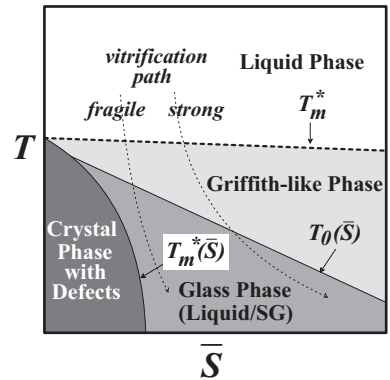,width=5cm}
\end{center}
\caption{Schematic phase diagram of the liquid-glass transition. 
$\bar{S}$ is a measure of disorder strength against density ordering. 
SG stands for spin-glass phase. Crystallization occurs around $T_m^\ast$ 
if it is not kinetically avoided.   
}
\label{fig:PD}
\end{figure}
\end{minipage}
\begin{minipage}{8cm}
\begin{figure}
\begin{center}
\psfig{figure=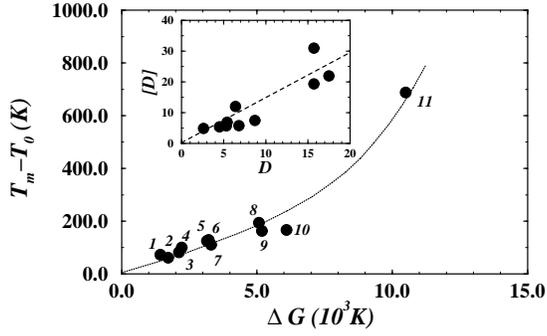,width=6cm,angle=-90}
\end{center}
\caption{Relation between $\Delta G$ and $T_m-T_0$. 
The data are taken from [12]. 
1: n-butyl benzene; 2: isopropyl benzene; 
3: propyl carbonate; 4: n-propanol; 5: o-terphenyl; 6: salol; 
7: dibutyl phtalate, 
8: s-triaphthyl benzene; 9: glycerol; 10: $\alpha$-phenyl-cresol; 
11: boron oxide. The curve is to guide the eye. The inset shows 
the relation between $[D]$ and $D$ (see the text on their definitions). }
\label{fig:TmT0}   
\end{figure}

\end{minipage}
\end{multicols}


\begin{references}
\bibitem{random} The randomness of the spatial distribution of locally 
favored structures is strongly supported by the absence of any 
spatially-correlated static 
heterogeneity in scattering experiments [1,2].
\bibitem{note} Note that $\delta \tau(\vec{r}) \leq 0$ in our case, 
since the bond formation always disturbs crystallization.
\bibitem{note2} Note that in some liquids like glycerol 
$\beta_K<1$ even above $T^\ast_m$. This may be due to 
the complexity  of molecules themselves 
(e.g., the distribution of hydrogen-bonging states).
\end{references}
\end{document}